\documentclass[twocolumn]{aastex631}

\shorttitle{2-D Cooling of Magnetic Neutron Stars}
\shortauthors{Negreiros et al.}
\graphicspath{{./}{figures/}}

\begin{document}

\title{Axisymmetric Cooling of Neutron Stars with Strong Magnetic Fields}
\newcommand{\vd}[1]{\textcolor{magenta}{#1}}
\newcommand{\jp}[1]{\textcolor{orange}{#1}}
\newcommand{\rn}[1]{\textcolor{blue}{#1}}
\newcommand{\msun}{$M/M_{\text{Sun}}$}

\author{R. Negreiros}
\affiliation{Department of Physics, Catholic Institute of Technology, MA, USA} 
\affiliation{Department of Physics, Universidade Federal Fluminense, Niteroi, Brazil}
\affiliation{ICRANet, Piazza della Repubblica 10, I-65122 Pescara, Italy}
\author{J. Peterson}
\affiliation{Department of Physics and Engineering, Western Wyoming Community College, Rock Springs WY 82901 USA}

\author{V. Dexheimer}
\affiliation{Department of Physics, Kent State University, Kent OH 44242 USA}

\begin{abstract}
We study the cooling evolution of neutron stars with strong poloidal magnetic fields (with strength not far from observed values) using the full general relativity 2-dimensional \textit{Astreus} code, which solves consistently Einstein's and Maxwell's equations. We find that central magnetic fields with strengths $3-4\times10^{17}$ G, corresponding to surface magnetic fields $7-8\times10^{16}$, can significantly modify the cooling behavior of neutron stars, leading to stars with similar masses but different magnetic fields to exhibit different thermal evolution. We show a non-linear increase in the thermal relaxation time with increasing magnetic fields and that this behavior is associated with the reduction of the Direct Urca process in stars with strong magnetic fields. This is a novel result in which we can observe the magnetic field influence on the thermal evolution of stars, even if it is not strong enough to affect the Fermi distribution of particles.
\end{abstract}


\section{Introduction}

Investigating the core of neutron stars presents a significant challenge. Beyond inferences derived from their observed masses and radii, few other observables are available to reveal insights into the interior of isolated objects \citep{MUSES:2023hyz}. In this study, we examine the cooling evolution of neutron stars endowed with magnetic fields powerful enough to disrupt spherical symmetry, thereby inducing deformations in their stellar structure. Specifically, we aim to explore how deviations from spherically symmetric geometry, driven by strong magnetic fields, might affect the thermal evolution of neutron stars.

Neutron stars are inherently non-spherical objects, exhibiting deformation due to rapid rotation and intense magnetic fields. Their formation results from asymmetrical supernova explosions \citep{cerda2019neutron,Ott_2018,burrows2013perspectives,janka2012explosion,Lander:2020bou}, and many undergo matter accretion from companion stars, leading to the development of accretion disks \citep{Hayasaki2004,Karino2019,PhysRevD.109.063023}. Consequently, it is meaningful to model neutron stars as axisymmetric structures when investigating their cooling evolution.

Neutron stars have been observed to exhibit surface magnetic fields of the order of $10^{8}$ to $10^{15}$ G. A subset of neutron stars, whose surface magnetic field may range from $10^{12}$ to $10^{15}$ G, are classified as magnetars \citep{Kaspi:2017fwg}. Whereas electromagnetic observation allows us to make reasonable estimates about the magnitude of magnetic fields in the surface of neutron stars, inferring the corresponding fields in the interior of neutron stars is a greater challenge; see, e.g., the estimation of $10^{16}$ G using slow phase modulations in hard X-ray pulsations \citep{Makishima:2014dua,Makishima:2018pmu,Makishima:2021vvv}. 

From a theoretical perspective, fully axisymmetric general relativistic solutions to Einstein's and Maxwell's equations suggest that the poloidal magnetic field in neutron stars intensifies as one moves from the surface toward the core \citep{Bocquet:1995je,Cardall:2000bs,Frieben:2012dz,Pili:2014npa,Dexheimer:2016yqu,Tsokaros:2021pkh}. This is in line with simple virial theorem estimates, which provide a rough upper limit for the internal magnetic field of neutron stars to be of the order of $10^{18}$ G \citep{1991ApJ...383..745L}. The same aforementioned solutions have demonstrated that stars whose surface and central fields reach $10^{16}$ G and $10^{17}$ G, respectively, become significantly deformed, thus requiring a fully axis-symmetric (2-dimensional) description \citep{Gomes:2019paw}. However, it must be noted that the field magnitude at which deformation can no longer be neglected depends on the composition of matter and interactions, as well as the stellar mass \citep{Rather:2022bmm}.

Even at the extreme densities found in neutron stars, their microscopic realm does not exhibit curvature. Therefore, the only requirements to describe dense matter in neutron stars are special relativity and quantum mechanics, which become the quantum field theory. Moreover, it is important to replicate the behavior of quantum chromodynamics (QCD) at high energies. This involves developing a theoretical framework that successfully captures both the restoration of chiral symmetry and the deconfinement transition to quark matter, expected to take place at high densities. In this study, we used the chiral mean field (CMF) model to achieve these objectives. It is important to note that within the scope of this work, the temperature is not microscopically significant. Following the proto-neutron star phase, which lasts approximately 60 seconds, the temperatures inside neutron stars drop well below the Fermi energy of their constituent particles. Consequently, all fermions adhere to a Fermi-Dirac distribution, rendering the temperature effects on the equation of state (EoS) and composition negligible. That is not to say that finite-temperature effects are always irrelevant, as they in fact become extremely important under certain circumstances, as in neutron star mergers, for instance. For a detailed discussion of the interaction of finite-temperature and magnetic-field effects in the CMF model, see~\citep{Peterson:2023bmr}.

Next, we describe the microscopic and macroscopic formalism used to describe dense matter under strong magnetic fields in this work in Sec.~2, with results for thermal properties shown in Sec.~3, and conclusions  presented in Sec.~5.

\section{Formalism}

{\bf Microscopic EoS}
To describe strongly interacting matter, we use the relativistic chiral mean field (CMF) model \citep{Dexheimer:2008ax}. It describes the baryon octet, nucleons and (strange) hyperons, interacting through mean-field mesons, allowing for strangeness to naturally appear as the density increases.  While the strong-force attraction is modeled by scalar mesons, the repulsion is modeled by vector mesons. Isospin asymmetry between, e.g., neutrons and protons is modeled by isovector mesons, and strangeness by mesons with hidden strangeness. A free (with respect to the strong force) gas of leptons (electrons and muons) is added to ensure charge neutrality. The beta equilibrium with leptons is enforced by determining the isospin fraction (equal to the lepton fraction) at a given density. 

In this work we do not include quarks in the EoS, as we focus on intermediate-mass stars that do not reach high densities in the core, although we do take into account chiral symmetry restoration (in the form of a decrease in the in-medium mass of the baryons). Furthermore, the effects on magnetic field in the microscopic composition and EoS are not considered,  as they only become relevant (in this context) above $10^{18}$ G, an extremely high value that is beyond the scope of this work. For the interested reader, please see \citep{Dexheimer:2011pz,Chatterjee:2014qsa,Franzon:2015sya,Peterson:2023bmr}

{\bf Macroscopic Structure}
In order to obtain results for macroscopic stellar properties, such as mass and radius, in the case of stars with strong magnetic fields we have to simultaneously numerically solve the Einstein and Maxwell equations.
We now briefly describe the structure equations for a general relativistic highly magnetic compact object. We begin by writing an axis-symmetric metric given by
\begin{eqnarray}\label{metric}
ds^2 &=& - e^{2 \nu} dt^2 + e^{2 \phi} (d\varphi - N^\varphi dt)^2
+ e^{2 \omega} (dr^2 + r^2 d\theta^2),\nonumber\\
\end{eqnarray}
with coordinates $x^{\mu}=(x^{0},x^{1},x^{2},x^{3})=(t,r,\theta,\varphi)$ and metric functions $\nu,\phi, \omega$ and $N^{\phi}$ that depend on coordinates $(r, \theta)$. The metric potentials are found by solving Einstein's equation coupled to Maxwell's equation in a curved space-time, 
\begin{eqnarray}\label{Einsteq}
G_{\mu \nu} = 8\pi T_{\mu \nu}, \ \ \ \rm{and}
 \ \ \ 
F^{\alpha \beta}_{\,\,\,\,\,\ ; \beta} = 4\pi j^{\alpha}. \label{eq:field_eq}
\end{eqnarray}
where $G_{\mu \nu}$ is Einstein's tensor, $T_{\mu \nu}$ the energy-momentum tensor, $j^{\alpha}$ the four-current, and the electromagnetic tensor $F_{\mu \nu}= A_{\nu, \mu} - A_{\mu, \nu}$,
where $A_\mu$ is the electromagnetic four potential. Here, comas and semi-colons denote ordinary and covariant derivatives, respectively.

The sources of curvature are given by the energy-momentum tensor, which is given by that of a perfect fluid ($PF$) \citep{Tolman:1939jz}. However, we must consider that we are interested in neutron stars whose magnetic fields are intense enough to cause curvature, and as such we must also include electromagnetic ($EM$) sources, which is achieved by including the stress-energy tensor for the electromagnetic field. We thus obtain 
\begin{eqnarray}\label{Tmunu}
T^{\mu \nu} &=& T^{PF \mu \nu} + T^{EM \mu \nu}
 \nonumber\\
 &=& (\epsilon + P)u^{\mu}u^{\nu} + P g^{\mu \nu} 
 \nonumber\\
 &+& \frac{1}{4\pi} \left( F^{\mu \alpha} F^{\nu}_{\,\,\, \alpha} - \frac{1}{4} g^{\mu \nu} F^{\alpha \beta} F_{\alpha \beta}  \right),
\end{eqnarray}
where $\epsilon$ is the energy density, $P$ the pressure, $u$ the fluid velocity, and $g$ the metric tensor.

Finally, by imposing the conservation of energy momentum ($T^{\mu \nu}_{\quad ;\nu} = 0$), we obtain the hydrostatic equilibrium equation for a relativistic, highly magnetic compact star, given by
\begin{eqnarray}
\frac{1}{(\epsilon + P)} P_{,i} + \nu_{, i} - (\ln \Gamma)_{,i} - \frac{1}{(\epsilon + P)}f_{i} = 0, \label{eq:hydro_eq}
\end{eqnarray}
where $\Gamma$ is the Lorentz factor. The third term in Eq.~(\ref{eq:hydro_eq}) results from the presence of an electromagnetic field (see, for example, \citep{cardall2001effects,bocquet1995rotating}) with the quantity $f_i$ denoting the Lorentz force, which is explicitly written as
$
f_{i} = F_{i \alpha} j^{\alpha} = j^{t}A_{t, i} + j^{\phi}A_{\phi, i}$, with distribution of charge $j^t$ and current $j^\varphi$.\\

{\bf Microscopic + Macroscopic Description}\\
Finally, to connect the microscopic realm with the macroscopic realm, we must also specify $j^t$ and $j^\varphi$. 
Considering that our purpose is to qualify the effects of magnetic field on the thermal properties of stars, we follow the standard approach \citep{cardall2001effects} and set $j^t = 0$, while adopting 
$
j^\varphi = f_0 (\epsilon + P), \label{eq:current}
$
with $f_0$ being a current function that is used to set the strength of the stellar magnetic field distribution. This choice leads to the formation of a purely poloidal magnetic field, which is what we desire to study. 

\begin{figure*}
    \includegraphics[width=0.52\textwidth]{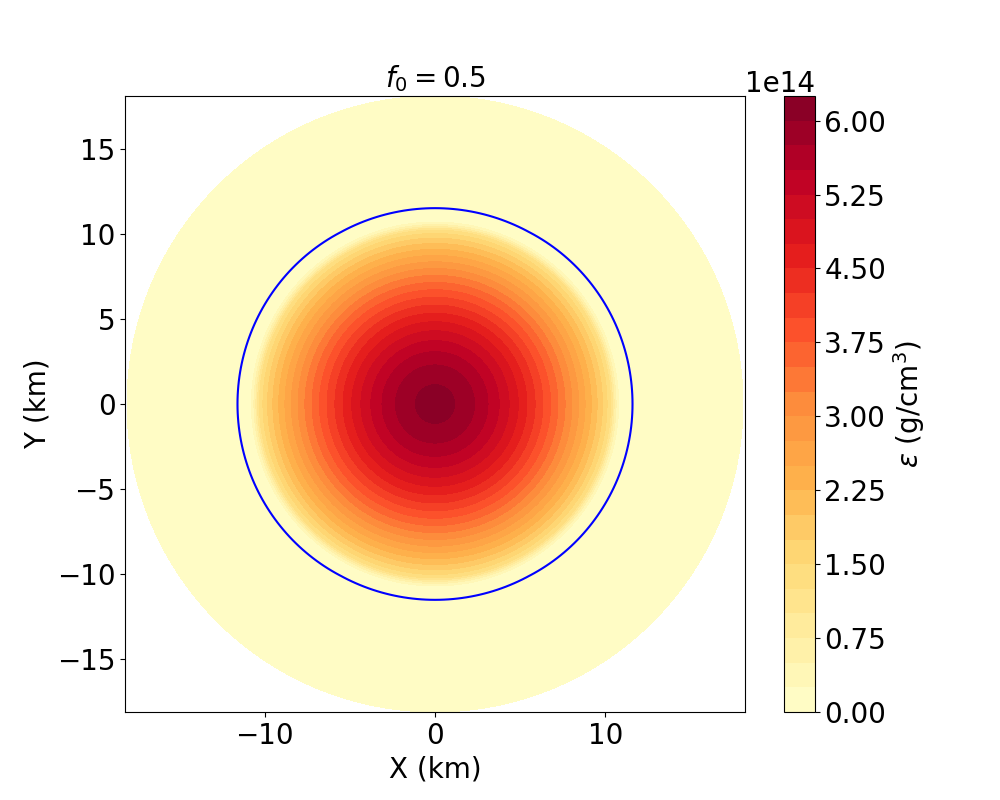}
    \includegraphics[width=0.52\textwidth]{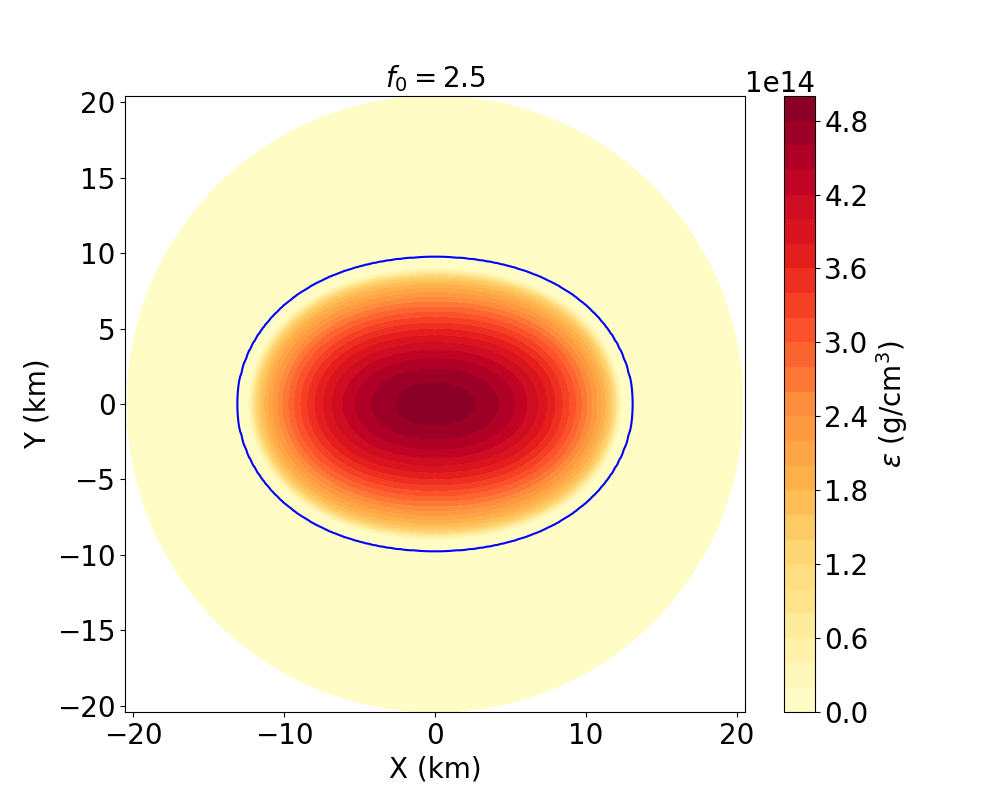}
    \includegraphics[width=0.52\textwidth]{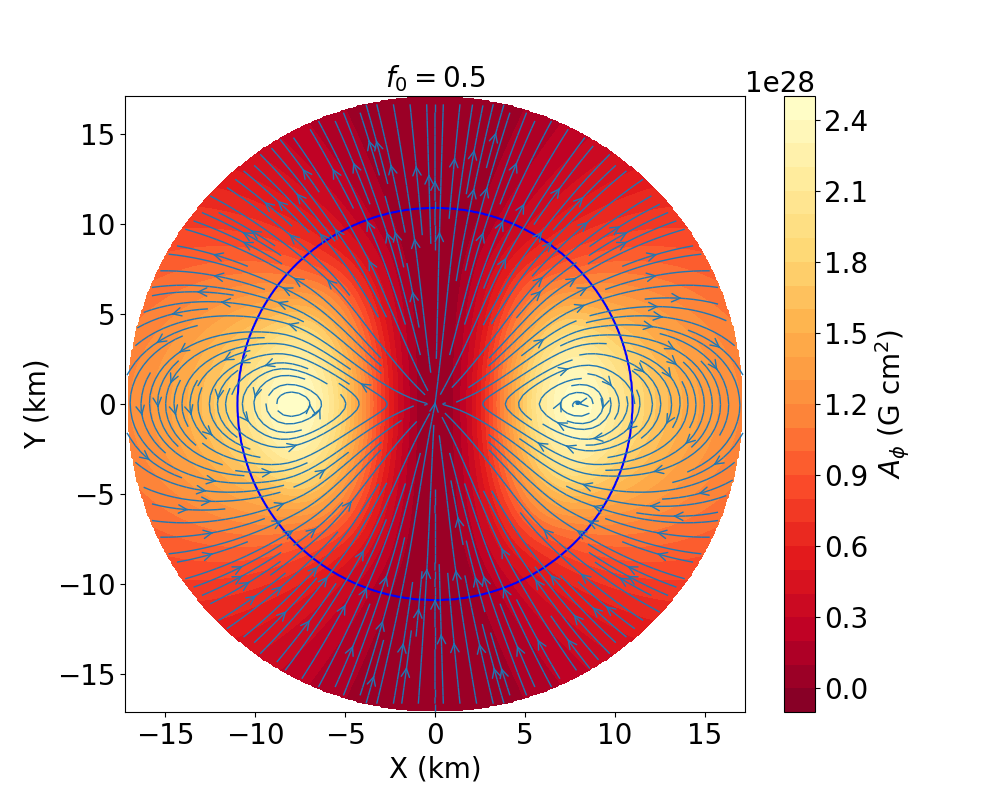}
    \includegraphics[width=0.52\textwidth]{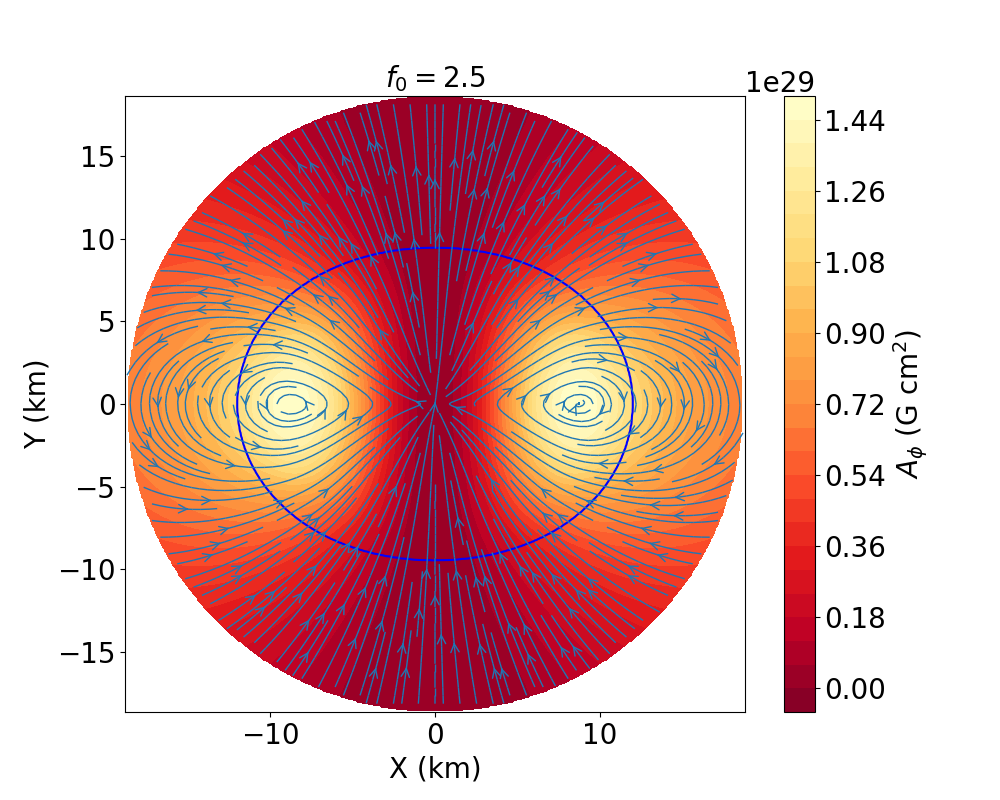}
  \caption{Top: Energy density contours for 1.4 \msun neutron stars with different current functions $f_0$. Bottom: Contours for the azimuthal component of the magnetic potential ($A_\phi$), as well as magnetic field lines, with different current functions $f_0$. The blue contours in all panels indicate the neutron star surfaces.}
      \label{fig:ED_map}
\end{figure*}

Within this approach, we have a closed description of the problem in which we have four metric potentials $\nu,\phi, \omega$ and $N^{\phi}$, one electromagnetic potential $A_3$, and two thermodynamic variables $P$ and $\epsilon$, totaling 7 variables. These are accompanied by four Einstein field equations Eq.~(\ref{eq:field_eq}), the hydrostatic equilibrium equation Eq.~(\ref{eq:hydro_eq}), the equation of state $P = P(\epsilon)$ and the current function Eq.~(\ref{eq:current}), totaling 7 equations, thus fully determining the problem.
These equations are solved numerically using the code \textit{Astreus}, developed by one of the authors and used previously to describe both rotating and highly magnetic compact objects \citep{peterson2021effects,negreiros2018many,negreiros2017fully,negreiros2012thermal}. The \textit{Astreus} code employs Green's function expansion to solve the field equations in the entire space, with the boundary conditions of a flat space at infinity. This numerical technique was originally developed by \citep{Komatsu1987} and expanded in \citep{Cook1992a}. 

In this study, we focus on neutron stars with a canonical mass of 1.4 solar masses. This choice was made because neutron stars with such mass can be readily obtained by numerous equations of state, in addition to the large number of observed objects (both electromagnetically \citep{Valentim:2011vs,Ozel:2012ax,Kiziltan:2013oja} and gravitationally \citep{Landry:2021hvl}). Our main focus with this study is to assess the impact of sufficiently strong, curvature-inducing, magnetic fields on the thermal behavior of neutron stars, rather than evaluating the accuracy of the underlying microscopic model in comparison to observed neutron star masses. 

We then numerically solve the structure equations and, by choosing different values for the current function $f_0$, we find several configurations for 1.4 solar mass stars, each with different magnetic field strengths. See Table~(\ref{tab:star_prop}) for magnetic field strengths at the center and pole for each current constant, together with different global properties of the corresponding stars. One can see, for instance, how the circumferential radius ($R_C$) of a star increases with magnetic field, growing from 13.83 km to 15.6 km when the central magnetic field reaches $4.7\times 10^{17}$ G. 

\begin{table}
    \centering
    \begin{tabular}{|c|c|c|c|c|}
    \hline
        $f_0$ & $R_c$ (km)  & $B_c$ ($10^{15}$G) & $B_p$ ($10^{15}$G) & $\mu$ 
        ($10^{34}$) \\
        \hline
        \hline
         0.0 & 13.83  & 0.0 & 0.0  & 0.0  \\
         \hline
         0.5 & 13.87  & 95.65  & 21.38  & 21.70 \\
         \hline
         1.0 & 13.96  & 191.89 & 41.44  & 44.52 \\
         \hline
         1.5 & 14.17  & 286.14 & 58.33  & 69.59 \\
         \hline
         2.0 & 14.50  & 383.89 & 72.45  & 100.73\\
         \hline
         2.5 & 15.16  & 473.64 & 80.54  & 144.35\\
         \hline
    \end{tabular}
    \caption{Different properties of $1.4$~\msun stars studied: circumferential radius $R_c$, central magnetic field $B_c$, polar magnetic field $B_p$, and magnetic moment $\mu$ (in Gaussians, where 1 Gaussian$=10^{-3}$ A.m$^2$).}
    \label{tab:star_prop}
\end{table}

To illustrate how the presence of  strong magnetic fields can change the geometry of stars (ultimately, we want to determine whether this has any effect on their thermal evolution), we illustrate the properties of two stars with current functions  $f_0 = 0.5$ and $f_0 = 2.5$. These values for $f_0$ represent low- and high-magnet-field configurations, with intermediate scenarios in between.
We begin by showing the energy density distribution inside the neutron stars in the top panels of Fig.~\ref{fig:ED_map}. The star with current function $f_0 = 0.5$ (left panel) has, effectively, a spherically symmetric geometry, even though it does have a non-vanishing magnetic field. The star with $f_0 = 2.5$ (right panel), on the other hand, exhibits a very deformed geometry, as is evident by the blue surface contour line (representing the stellar surface). The difference in geometry exhibited by these two stars is associated with the magnitude of the magnetic field in these stars, with the field in the $f_0 = 0.5$, albeit large, not being strong enough to cause significant curvature. Unlike the star with $f_0 = 2.5$ has magnetic fields sufficiently intense to alter space-time, imprinting their symmetry to it, thus leading to a deformed, axis-symmetric star.

The bottom panels of Fig.~\ref{fig:ED_map} clearly show the dipolar configuration of the magnetic field, which is qualitatively the same for both stars. However, the central magnetic field for $f_0 = 2.5$ (right panel) is approximately 7 times higher than for $f_0 = 0.5$ (left panel), and this is sufficient to cause extreme deformation in the star. This behavior, which has been observed in previous work, elucidates the non-linear nature of the magnetic field influence, with curvature effects rapidly becoming relevant as soon as the field reaches a certain value $\sim 10^{17}G$ \citep{Gomes:2019paw}.
\section{Thermal Properties}

We now investigate the thermal properties of the stars discussed in the previous section by solving the thermal evolution equation of neutron stars in a fully axis-symmetric space-time, given by the following parabolic equation \citep{negreiros2012thermal,negreiros2017fully}
\begin{eqnarray}
\partial_r \tilde H_{\bar r} + {1 \over r} \partial_\theta \tilde
H_{\bar\theta} &=& - r \, e^{\phi + 2(\alpha-\beta)} \left( {e^{2\nu}
  \over \Gamma} \epsilon + \Gamma C_V \partial_t \tilde T\right),\nonumber\\
\partial_r \tilde T &=& - {1 \over{r \kappa}} e^{-\nu -\phi} \tilde H_{\bar r},
\,
\nonumber\\
{1 \over r} \partial_\theta \tilde T &=& - {1 \over{r \kappa}} e^{-\nu
  -\phi} \tilde H_{\bar \theta}, \,
\quad \quad \quad
\label{eq:Thermal_eq} 
\end{eqnarray} 
where $\alpha - \beta = \omega$ (metric), $\tilde H_i \equiv r e^{2\nu+\phi+\omega} H_i / \Gamma$, with $H_i$ being the i-th component of the heat flux; $\tilde T \equiv
e^\nu T / \Gamma$, with $T$ being the temperature; $\kappa$ is the
thermal conductivity; $C_V$ is the specific heat; $\epsilon$ is the
neutrino emissivity; and the Lorentz factor $\Gamma \equiv (1 -
U^2)^{-1/2}$, where $U$ is the proper velocity with respect to a
zero angular momentum observer, given by $U = (\Omega -
N^\varphi)e^{\phi}$ with $\Omega=0$ for the current study. 

\begin{figure}
    \centering
    \includegraphics[width=\linewidth]{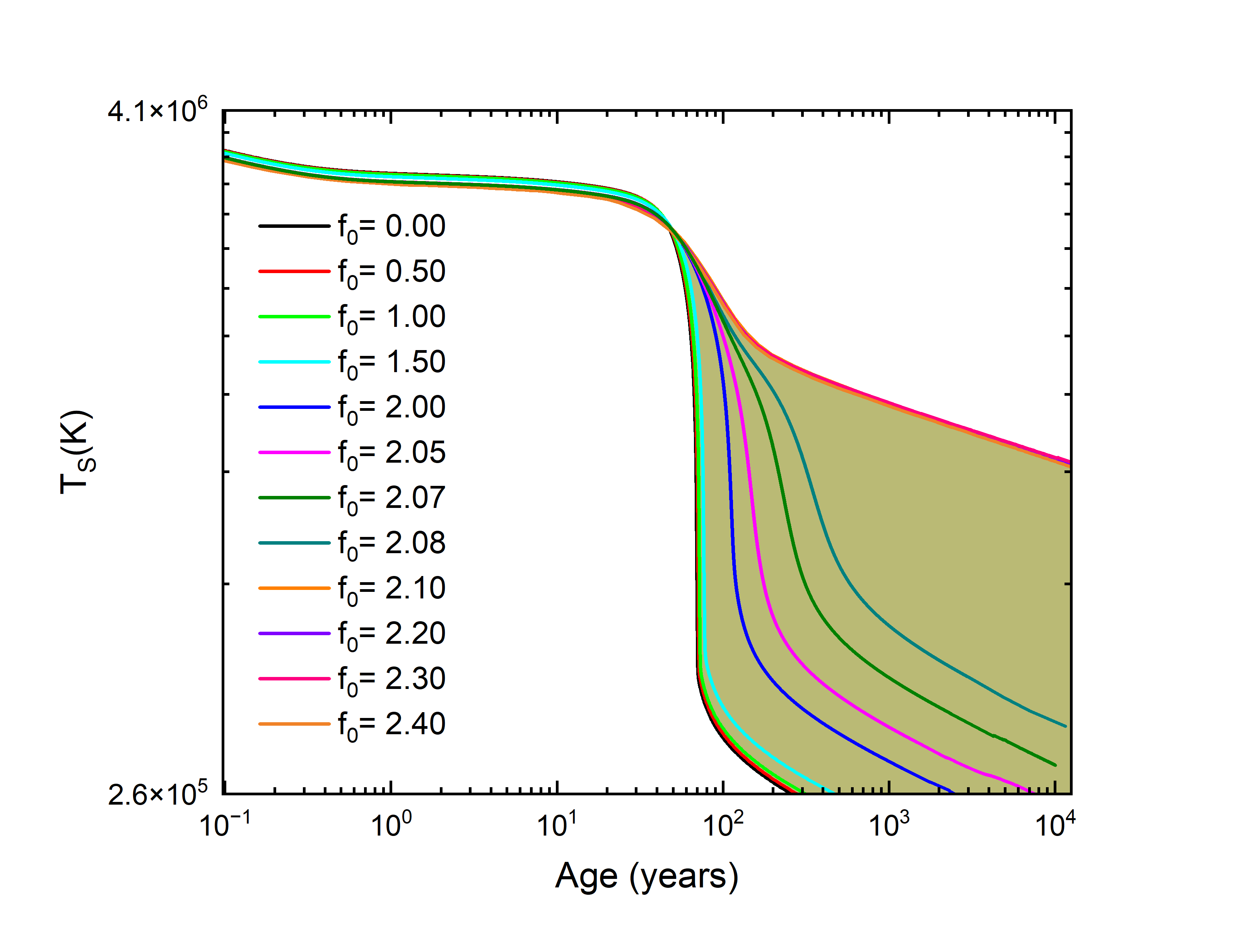}
    \caption{Redshifted temperature evolution for the equatorial region of stars with $1.4$~\msun and different values of the current function $f_0$. The olive shaded region denotes possible different evolutions for current functions comprised between $f_0 = 2.0$ and 2.5.}
    \label{fig:T_eq}
\end{figure}

In this work, we follow the ground work laid out
in \citep{negreiros2012thermal}, except here we consider neutron stars deformed by strong magnetic fields, and proceed to numerically solve the thermal evolution
equations Eq.~(\ref{eq:Thermal_eq}). We employ an Alternating Direction Implicit (ADI) method to integrate the cooling equations. Furthermore, we consider all
neutrino emission processes that may occur inside the neutron star,
including direct and modified Urca processes, Bremsstrahlung, and pair
breaking/formation. A review of such processes may be found in
\citep{Yakovlev2004,Page2004,2006NuPhA.777..497P}.

We show in Fig.~\ref{fig:T_eq} the evolution of the red-shifted surface temperature for the equatorial region of the stars in  Table~(\ref{tab:star_prop}). It is very intriguing that stars of identical mass can display a broad range of cooling behaviors depending on their magnetic fields. 
Here, we note that our microscopic model allows for the direct Urca (DU) process in stars with 1.4 \msun, which explains the fast cooling exhibited by spherically symmetric stars. This is usually alleviated by the inclusion of appropriate pairing among nucleons. We will conduct a thorough analysis including a sophisticated pairing scheme in a future work, once we have a comprehensive understanding of the cooling of stars with strong magnetic fields. 

\begin{figure}
    \centering
    \includegraphics[width=\linewidth]{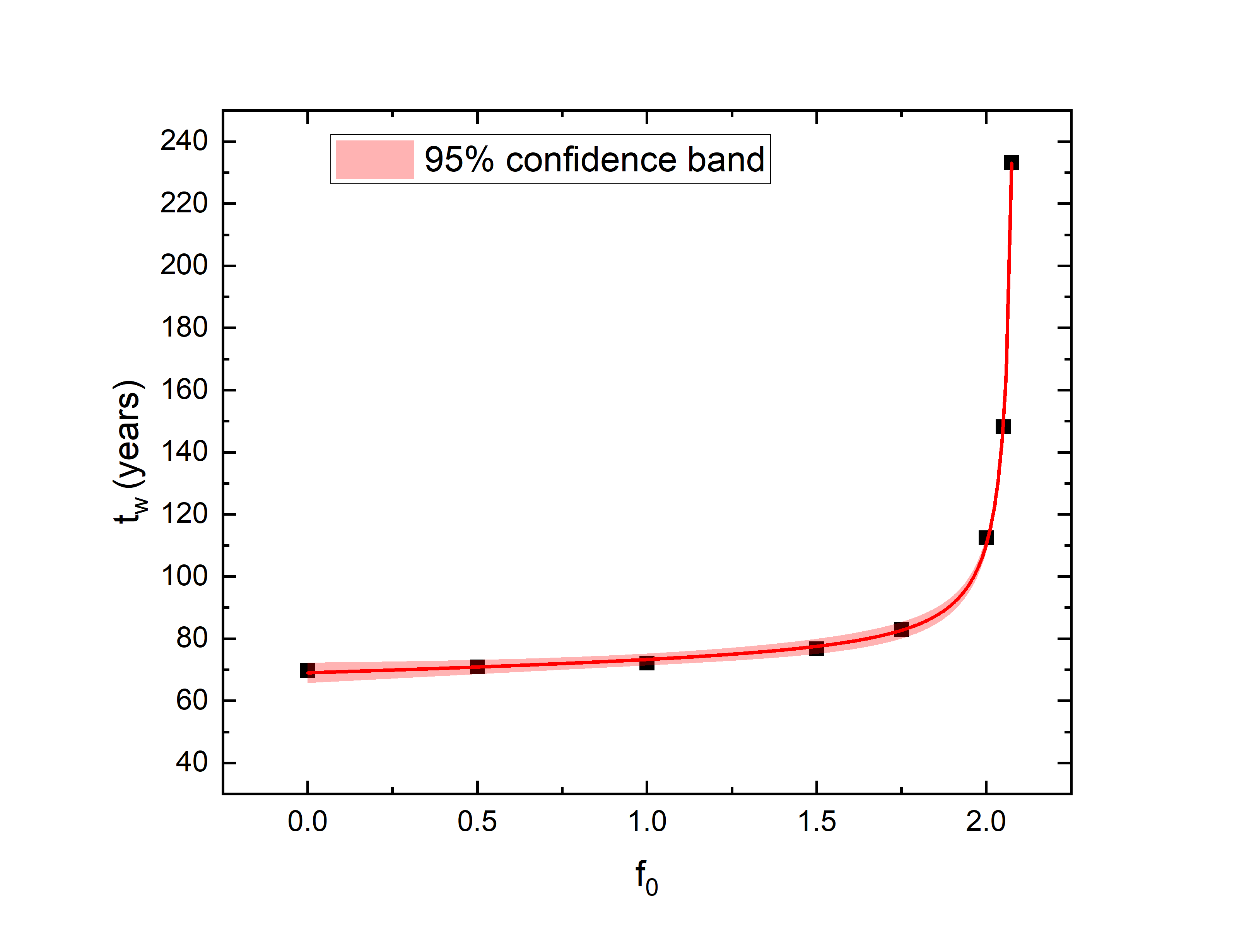}
    \caption{Relaxation time as a function of current parameter $f_0$. The red shaded region represents the 95\% confidence band for the curve fitted to the data.}
    \label{fig:twxf0}
\end{figure}

Regardless of pairing or lack thereof, the results shown in Fig.~\ref{fig:T_eq} indicate that stars with relatively low magnetic fields, such as those with $0.0 \leq f_0 \leq 1.5$ have a thermal evolution very similar to their spherically symmetric counter-part ($f_0 = 0$), that is, a fast cooling with a sharp drop at ages $\sim 100$ years, which is characteristic of stars with the DU taking place \citep{sales2020revisiting,zapata2022thermal}. We can also see, however, that the stars with higher current functions, and thus higher magnetic fields,  exhibit higher surface temperatures. Interestingly if $f_0$ becomes high enough we start to have stars that exhibit slow cooling, typical of stars without the DU process. The olive-shaded region in Fig.~\ref{fig:T_eq} indicates the range of different cooling curves that appear for a continuous variation of $f_0$ between 2.0 and 2.5. 

\begin{figure*}
    \includegraphics[trim={3cm 1.5cm 0 0.0cm},clip,width=0.52\textwidth]{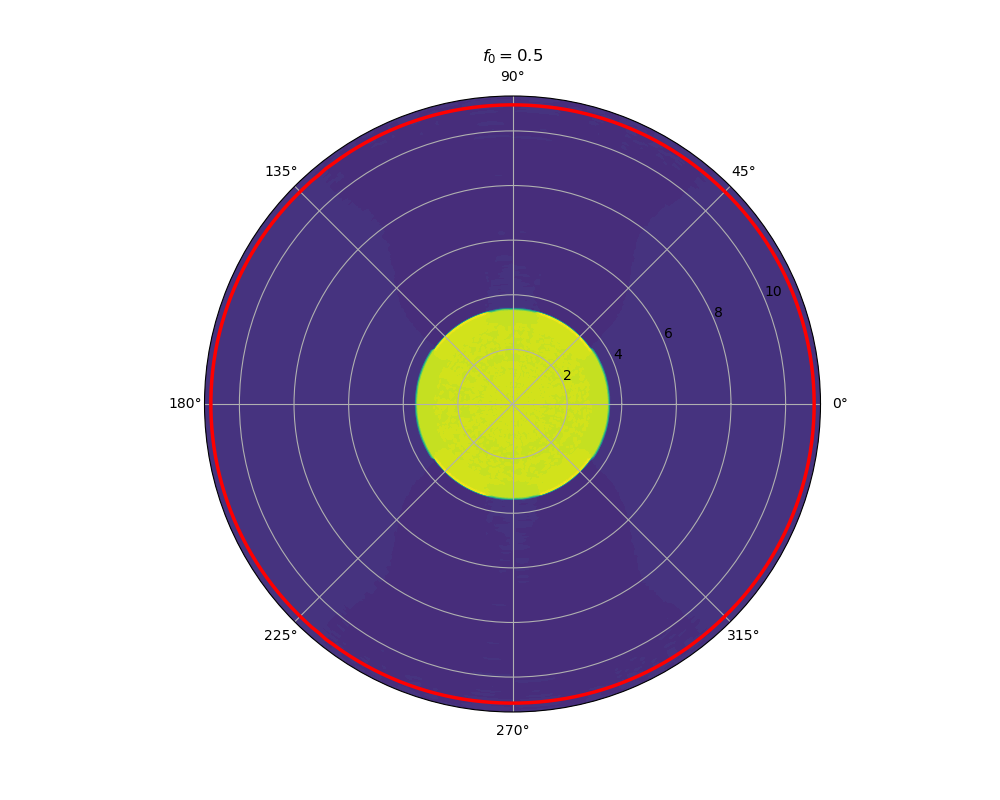}
    \includegraphics[trim={3cm 1.5cm 0 0cm},clip,width=0.52\textwidth]{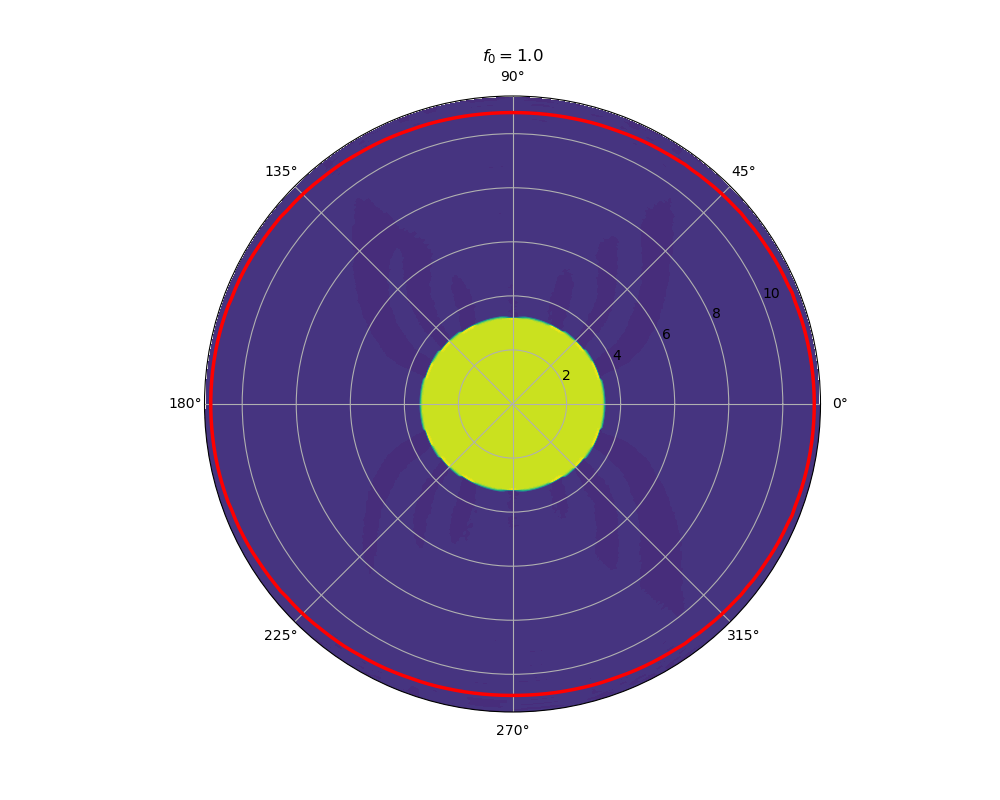}
    \includegraphics[trim={3cm 1.5cm 0 0cm},clip,width=0.52\textwidth]{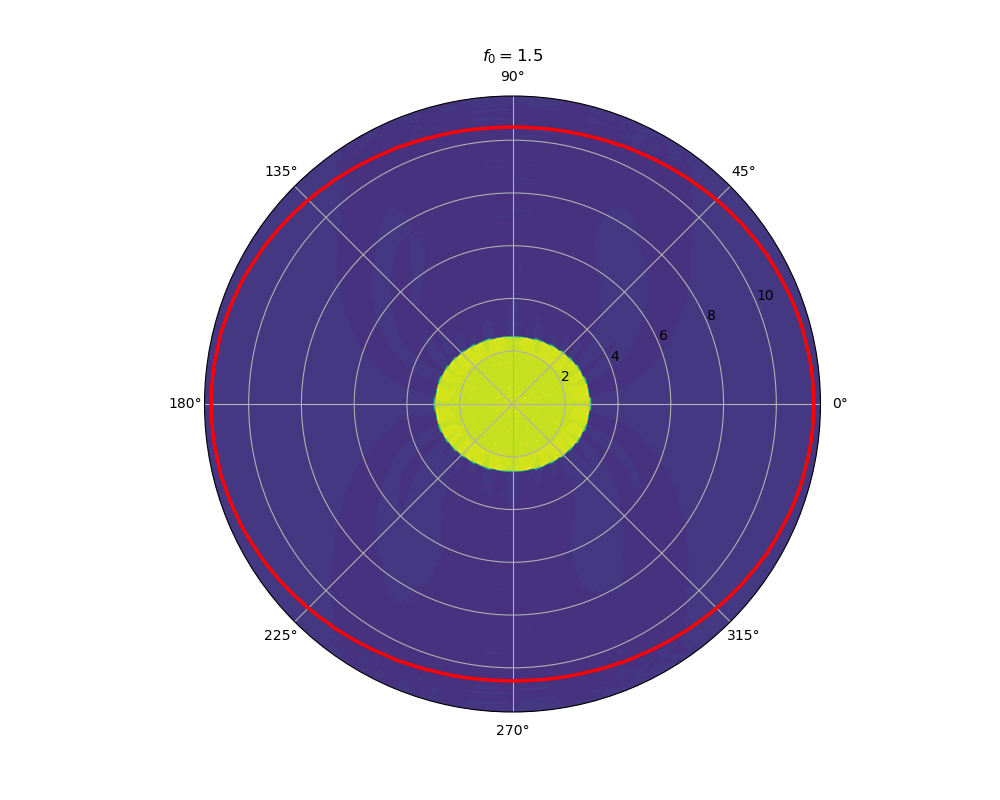}
    \includegraphics[trim={3cm 1.5cm 0 0cm},clip,width=0.52\textwidth]{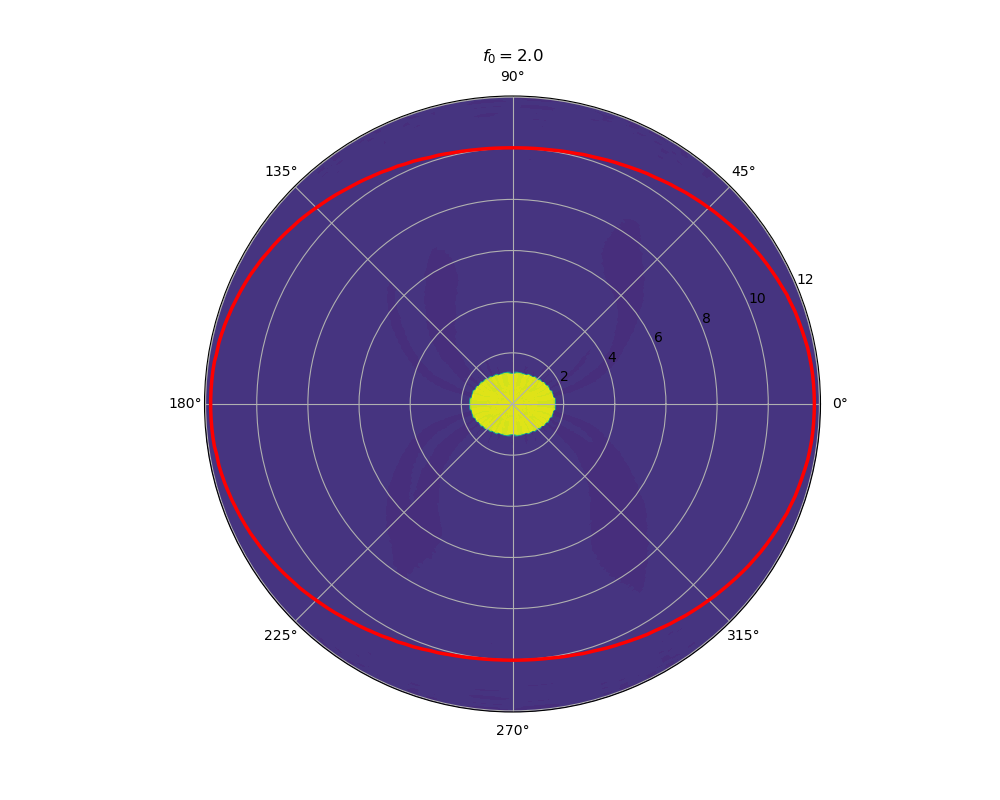}
  \caption{Active regions for the Direct Urca process (yellow shaded region) inside of 1.4 solar mass stars with different magnetic fields distribution, according to different values of the current constant $f_0$.  Red lines indicate the stellar surface.}
      \label{fig:DU_map}
\end{figure*}

These results may appear surprising, as it has already been shown that magnetic fields of magnitude considered in this paper ($\sim 10^{17}$ G) are not high enough to alter the equation of state of neutron star matter \citep{Chatterjee:2014qsa,Franzon:2015sya}, which would naively indicate that they could not affect the thermal evolution of the star. However, our results show that this is not the case. Even though the magnetic fields cannot alter the equation of state, they are strong enough to give rise to curvature, and thus contribute to the gravitational mass observed by an observer at infinity. This means that stars with the same gravitational mass ($1.4$~\msun in the case studied) but different magnetic fields must have different baryonic contents, as the total gravitational mass is a result of the curvature caused by the baryonic and electromagnetic content. Thus, when comparing two stars with the same gravitation mass, a highly magnetic object has a lower baryonic content, and thus a lower baryonic density, which in turn reduces the region in which the DU process is active, as the DU needs a certain proton fraction to be permitted. 

This result is reflected in the relaxation time of neutron stars. This quantity is defined as 
\begin{equation}
    t_w = \max \left| \frac{d\ln T_s}{d\ln t}\right|,
\end{equation}
and represents the time in which the core-and the crust become thermally coupled. This quantity, which is connected to both the volume of the core in which the DU takes place \citep{sales2020revisiting} and the thickness of the crust, would naturally be influenced by a neutron star with a magnetized structure. As discussed above, the increase in the magnetic field strength changes the volume in which the DU takes place. It also affects the thickness of the crust, as was discussed in \citep{franzon2017magnetic}. 
The resulting relaxation time, as a function of the current function, is shown in Fig.~\ref{fig:twxf0}. 
It shows a significant increase in the relaxation time as $f_0$ grows. After this constant reaches a value of approximately 2.0, we see a change of the thermal evolution regime from fast to slow cooling. We also note that, as shown in Fig.~\ref{fig:twxf0}, a good fit to the curve
\begin{equation}
    t_w = \frac{a - bf_0}{1 + cf_0 +df_0^2},
\end{equation}
is achieved, with the parameters being: $a =69.2, b=-32.13, c =-0.51$ and $d =0.18$.

As pointed out in Ref.~\citep{sales2020revisiting}, a non-linear increase in the relaxation time is typically associated with the transition from a fast to a slow cooling regime. This is also the case here, with the reduction in the central density of the stars, caused by the increase in magnetic field strengths, leading to a reduction in the volume in which the Direct Urca Process is active in the stellar core. This is confirmed by the results outlined in Fig.~\ref{fig:DU_map}.
They shed light on the behavior of the relaxation time discussed previously: One can see that as the magnetic field increases, the region in which the DU is active within the star is significantly reduced. Furthermore, it also loses its spherical shape, as the whole geometry of the star moves into an ellipsoidal geometry. This is a novel result in which we can observe how magnetic fields influence the thermal evolution of stars, even if they are not strong enough to affect the Fermi distribution of the particles. 

\section{Conclusions}

In this work, we describe the thermal evolution of highly magnetic neutron stars. To describe the microscopic equation of state (EoS), we include the whole baryon octet (nucleons and hyperons) and describe the strong interaction using the CMF model. We construct the structure of beta-equilibrated charge-neutral stars with strong poloidal magnetic fields (with strength not far from observed values) using the full general relativity 2-dimensional \textit{Astreus} code, which solves Einstein's and Maxwell's equations.  We then study the thermal evolution of axis-symmetric stars by solving thermal evolution equations. 

We focus on canonical stars that have masses of $1.4$~\msun. We start by allowing for all relevant neutrino emission processes. Then, we find that for stars with central magnetic fields with strengths $3-4\times10^{17}$ G, corresponding to surface magnetic fields of
$7-8\times10^{16}$ G, a strong magnetic field can significantly modify the cooling evolution of neutron stars, allowing stars to remain hotter by not fulfilling the conditions for the direct Urca process to take place.

In the future, we will extend our work to also describe the thermal evolution of highly magnetic proto-neutron stars. This extension of our work could be very interesting, as there have already been hints that the 2-dimensional neutrino distribution on proto-neutron stars is very sensitive to the magnetic field strength, as well as temperature \citep{Franzon:2016iai}. We emphasize our objective of integrating novel effects into our two-dimensional approach, particularly those by which magnetic fields can influence the thermal evolution of neutron stars. These include axion degrees of freedom, internal heating mechanisms, and superconductivity \citep{yadav2024thermal,10.1093/mnras/stab3126,sinha2015magnetar}. Additionally, future research should investigate the possibility of pairing, as deviations from spherical symmetry are expected to modify the geometry of the superfluid phase within neutron stars. Such alterations will consequently impact both the relaxation time and the resulting anisotropic temperature distribution across stars.

\section*{Acknowledgments}

We acknowledges support from the Department of Energy under grant DE-SC0024700 and from the National Science Foundation under grants MUSES OAC-2103680 and NP3M PHY2116686.

\bibliography{bibliography}{}
\bibliographystyle{aasjournal}

\end{document}